\documentclass[prl,preprint,showpacs,amsmath,amssymb]{revtex4}
\usepackage{graphicx}
\usepackage{dcolumn}
\usepackage{bm}
\begin{document}

\title{Relating breakup and incomplete fusion of weakly-bound nuclei 
through a classical trajectory model with stochastic breakup}

\author{A. Diaz-Torres$^1$, D.J. Hinde$^1$, J.A. Tostevin$^{1,2}$, M.
Dasgupta$^1$ and L.R. Gasques$^1$}

\affiliation{$^1$Department of Nuclear Physics, Research School of
Physical Sciences and Engineering, Australian National University,
Canberra, ACT 0200, Australia \\
$^2$Department of Physics, University of Surrey, Guildford, GU2 7XH,
United Kingdom}
\date{\today}

\begin{abstract}
A classical dynamical model that treats break-up stochastically is 
presented for low energy reactions of weakly-bound nuclei. The 
three-dimensional model allows a consistent calculation 
of breakup, incomplete and complete fusion cross sections. The model 
is assessed by comparing the breakup observables with 
CDCC quantum mechanical predictions, which are found to be in reasonable 
agreement. Through the model, it is demonstrated that the breakup probability 
of the projectile as a function of its distance from the target is of primary 
importance for understanding 
complete and incomplete fusion at energies near the Coulomb barrier.
\end{abstract}


\maketitle

Recent developments of radioactive isotope accelerators provide
an opportunity to investigate on Earth the fusion reactions that form heavy
elements in the cosmos. These involve reactions of nuclei far from
stability, the most exotic of which are often very weakly bound.
Breakup of weakly bound nuclei is thus an important process in their collisions 
with other nuclei. A major consequence of breakup is that not all the
resulting breakup fragments might be captured by the target, termed
incomplete fusion ({\sc icf}); capture of the entire projectile by the target 
is called complete fusion ({\sc cf}). 
Such {\sc icf} processes can dramatically change the nature of the reaction
products, as has been investigated in detail for the stable
weakly-bound nuclei $^9$Be and $^{6,7}$Li \cite{Dasgupta0}. There,
at energies above the fusion barrier, {\sc cf} yields were found to be only 
$\sim 2/3$ of those expected, the remaining $1/3$ being in {\sc icf} products. 
Events where the projectile breaks up
and none of the fragments are captured provide an important diagnostic 
of the reaction dynamics. This we call no-capture breakup ({\sc ncbu}), 
also referred to as elastic breakup.

In a conventional picture of fusion, two colliding nuclei will
fuse if they overcome the potential barrier due to their mutual
Coulomb and nuclear interactions. The additional breakup degrees of
freedom when one of the colliding nuclei is weakly-bound makes the
process very much more complicated. An outstanding theoretical challenge is 
to model the {\sc cf} and {\sc icf} processes in such
collisions, since this separation is crucial to understand the
effects of breakup on fusion \cite{Dasgupta0,Gomes0}. Quantum
mechanical few-body approaches, such as the continuum discretised 
coupled-channels (CDCC) method
\cite{Hagino0,Alexis0} and the time-dependent wave packet
method \cite{Yabana}, cannot separate incomplete and complete fusion
contributions to their absorptive cross sections \cite{Ian0}, since both 
result in depletion of the total few-body wave function. The CDCC method 
can, however, make reliable predictions of the {\sc ncbu} 
process \cite{TostevinBU}, as will be exploited here. 
What then are the alternatives to the above models? A novel
optical decoherence model has been suggested \cite{Ian0} but has yet
to be implemented. Another approach is to use the concept of classical trajectories 
which allow {\sc cf} and {\sc icf} events to be separated, as in the 
two-dimensional model of Ref. \cite{Hagino2}. Classical dynamical models are also 
being successfully applied to describe atomic many-body collisions, including 
ionization \cite{Arbo}, a process in some ways analogous to breakup.

In this Letter we present a new three-dimensional classical
dynamical reaction model that allows quantitative calculations of {\sc cf} and {\sc icf} 
yields above the barrier. This is achieved by the introduction of a stochastically 
sampled breakup function. 
Crucially, this function can be determined from experimental sub-barrier 
breakup data \cite{HindeBU}, or from CDCC calculations. 
We apply the model to two-body projectiles,
although it can be extended to more complex many-body breakup
scenarios. As individual trajectories are followed, the model allows
a clear identification of {\sc cf}, {\sc icf} and {\sc ncbu} events,
allowing predictions of their yields as well as breakup fragment
angular, kinetic energy and relative energy distributions. This
unique capability provides a powerful tool for
interpreting future fusion experiments with radioactive nuclei, and
in applications to $\gamma$-ray spectroscopy \cite{Dracoulis}. 
The model is assessed quantitatively by comparing its 
predictions for {\sc ncbu} observables with those of fully quantum mechanical 
CDCC calculations. These are carried 
out for a simplified test case. The classical model {\sc cf} and {\sc icf} 
predictions are then compared to experimental data.

The classical model calculations are as follows. The weakly bound (two-body) 
projectile $P$, with incident energy $E_0$ and orbital angular momentum $L_0$, is
incident on the target $T$, initially at rest in the laboratory
frame. Prior to breakup the projectile follows a classical orbit
with a $P$-$T$ distance of closest approach $R_{min}(E_0,L_0)$.
These orbits of the bound projectile are calculated by numerical solution
of the classical equations of motion in the presence of the Coulomb and nuclear 
potentials between $P$ and $T$. These generate a Coulomb barrier for head-on
($L_0=0$) collisions of height $V_{B}^{PT}$ at a separation
$R_{B}^{PT}$ .

The effects of the Coulomb and nuclear two-body interactions that cause breakup are 
encoded in a local 
breakup probability ${\cal P}^L_{BU} (R)$, a function of the
projectile-target separation $R$, such that ${\cal P}^L_{BU}(R) dR$ is the probability 
of breakup on the interval $R$ to $R+dR$.  
Crucially, for a given projectile-target combination, 
both experimental data \cite{HindeBU} and
CDCC calculations indicate that the integral
of this breakup probability along a given classical orbit is an
exponential function of its distance of closest approach,
$R_{min}(E_0,L_0)$. Explicitly,
\begin{equation}
{P}_{BU}(R_{min})=2\int_{R_{min}}^\infty {\cal P}^L_{BU}(R) dR =
A\,\exp (-\alpha R_{min}), \label{one}
\end{equation}
from which it follows uniquely that the local function has the same
exponential form, ${\cal P}^L_{BU}(R) \propto \exp (-\alpha R)$. 
  
For each $L_0$ (chosen to be an integer number of $\hbar$) a sample
of $N$ incident projectiles is taken. The
position of breakup on this orbit is determined by sampling a breakup
radius $R_{BU}$ on the interval $[R_{min}(E_0,L_0),\infty]$ with the
weighting ${\cal P}^L_{BU}(R)$. Although dependent on the constants
$A$ and $\alpha$, the exponential weighting of ${\cal P}^L_{BU}(R)$
will clearly place most $R_{BU}$ in the vicinity of $R_{min}$. 
If the chosen $L_0$ is less than the critical partial wave for 
projectile fusion, $L_{cr}$, then the associated
trajectory would normally lead to {\sc cf}, i.e. $R_{min}\leq
R_{B}^{PT}$. For these $L_0$ we set $R_{min} = R_{B}^{PT}$, 
when sampling $R_{BU}$, and all
breakup events are confined to the incoming branch of the
projectile trajectory. For $L_0 > L_{cr}$ breakup can take place on
both the entrance and the exit branches of the classical orbit, which
are sampled equally. Having chosen the position with the 
projectile center of mass at breakup, it is instantaneously broken 
up into fragments F1 and F2. 

Following breakup, fragments F1 and F2 now interact with $T$, 
and with each other, via
real central two-body potentials having Coulomb barriers
$V_{B}^{ij}$ at separations $R_{B}^{ij}$, $i,j=1,2,T,\ i\neq j$. 
The instantaneous dynamical variables of the excited projectile at
breakup, its total internal energy $\varepsilon_{12}$, angular
momentum $\vec{\ell}_{12}$ and the separation of the fragments
$\vec{d}_{12}$ are all Monte Carlo sampled. The
initial separations $d_{12}$ between the fragments are Gaussian
distributed in their classically allowed region to mimic the radial
probability distribution of the projectile ground-state (g.s.) wave
function. For high ${\ell}_{12}$ excitations, when there is no barrier 
between F1 and F2, $d_{12}$ is equated with their external turning point. 
In the test calculation below, the orientation of 
$\vec{d}_{12}$ is chosen 
randomly over the $4\pi$ solid angle, whilst the orientation of 
$\vec{\ell}_{12}$ is chosen randomly from all directions orthogonal to
$\vec{d}_{12}$. Other distributions could easily be encoded in cases 
where the two breakup fragments are not identical. ${\ell}_{12}$ is 
sampled uniformly on the interval
[0,$\ell_{max}$]. For $\varepsilon_{12}$ we tested two sampling
functions for energies between the top of the barrier,
$V_{B}^{12}$, and a chosen maximum $\varepsilon_{max}$. 
Using uniform weighting or an exponentially decreasing weighting 
yielded very similar outcomes, although the convergence of the
observables was faster for the latter, which was therefore used. 
Both $\ell_{max}$ and $\varepsilon_{max}$ were increased until 
convergence of the observables occured, as in the CDCC calculations.

Having fixed the positions and dynamical variables of the excited projectile 
fragments at the moment of breakup, the instantaneous velocities of the particles
F1, F2 and $T$ are dictated by conservation of energy, linear
momentum and angular momentum in the overall center of mass frame of
the projectile and target system. These breakup initial
conditions were transformed to the laboratory frame where the
equations of motion are solved. The
computed trajectories of F1, F2 and $T$ determine the number of {\sc
icf}, {\sc cf} and {\sc ncbu} events, fragment F$j$ being assumed to
be captured if the classical trajectories take it within the 
fragment-target barrier radius $R_{B}^{jT}$.

From the $N$ breakup events sampled for each projectile 
angular momentum $L_0$, the numbers of events $N_i$ in which $i=$ 0
({\sc ncbu}), 1 ({\sc icf}), or 2 ({\sc cf}) fragments are captured
determine the relative yields $\widetilde{P_i}=N_i/N$ of these
three processes after breakup, with $\widetilde{P_{0}} +
\widetilde{P_{1}} + \widetilde{P_{2}} = 1$. The absolute probabilities
$P_i(E_0,L_0)$ of the three reaction processes are determined from the 
relative yields and the integrated breakup probability over the whole 
trajectory $P_{BU}(R_{min}$):
\begin{eqnarray}
P_{0}(E_0,L_0) &=& P_{BU}(R_{min})\,\widetilde{P_{0}},\label{eq5} \\
P_{1}(E_0,L_0) &=& P_{BU}(R_{min})\,\widetilde{P_{1}},\label{eq3} \\
P_{2}(E_0,L_0) &=& [1-P_{BU}(R_{min})]\,H(L_{cr}-L_0) \nonumber\\
               &+& P_{BU}(R_{min})\, \widetilde{P_{2}},
\label{eq4}
\end{eqnarray}
where $H(x)$, the Heaviside step function, is unity for $x \geq 0$.
Finally, the physical cross sections are calculated
using
\begin{equation}
\sigma_i (E_0)= \pi \lambdabar^2 \sum_{L_0} (2L_0 + 1) P_i(E_0,L_0)
\end{equation}
where $\lambdabar^2=\hbar^2/[2m_P E_0]$ and $m_P$ is the projectile
mass. Other observables, e.g. the angle, kinetic energy and
relative energy distributions of the fragments from {\sc ncbu} 
events, are calculated by tracking their trajectories to a large 
distance from the target (200 fm in the calculations 
presented below).

In order to validate our approach, {\sc ncbu} observables, reliably 
calculated within the CDCC quantum mechanical approach, are compared with the 
classical model calculation. We do so in a simplified model 
case for the reaction of a pseudo-$^{8}$Be projectile $P$ (assuming a
weakly-bound state of two $\alpha$-particles) with a $^{208}$Pb target. 
The CDCC and classical model calculations are matched as closely as possible, 
as discussed below.

We bind $P$ in a nodeless $s$-state with a separation energy of 0.5
MeV in the Woods-Saxon (WS) potential well ($V$, $r$, $a$) $\equiv$
(16.7 MeV, 1.20 fm, 0.62 fm). The same potential was used to
construct the continuum states of $P$, which are non-resonant in all
partial waves. The Coulomb potential between all particles is that of 
uniformly charged spheres with a radius parameter 1.2 fm. The CDCC 
breakup model space includes (i) even $\ell_{12}$ up to 4$\hbar$, 
(ii) a maximum $\alpha$--$\alpha$ relative energy of 12 MeV, 
divided into 12 bins in each partial wave, and (iii) $\alpha$--$T$ 
potential multipoles 
up to quadrupole terms. The $\alpha$-$T$ optical potential is the 
real WS interaction (62 MeV, 1.39 fm, 0.62 fm). A diagonal absorptive 
WS potential (10 MeV, 1.30 fm, 0.74 fm) is included in all channels, 
its range chosen so that absorption occurs inside the fusion radius 
of the $\alpha$-$T$ system, $R_{B}^{\alpha T} = 10.6$ fm. 
Calculations were carried out at a laboratory energy of $E_0$ = 45 MeV.

\begin{figure}
\includegraphics[width=7.0cm,angle=0]{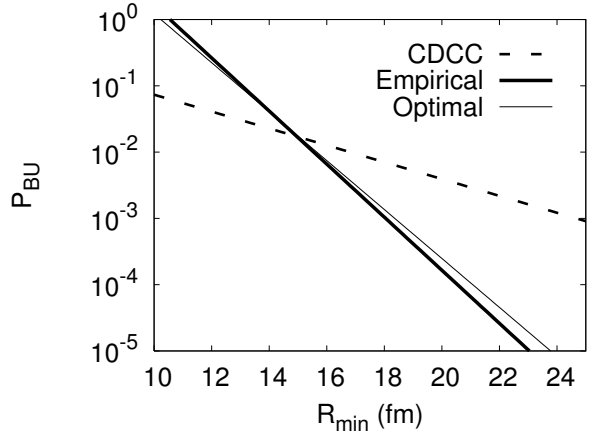}
\caption{The $P_{BU}$ versus $R_{min}$ from the CDCC model, 
the empirical line from the $^{9}$Be+$^{208}$Pb sub-barrier data 
\protect\cite{HindeBU}, and the optimal function 
used in Fig.\ \protect\ref{fig4}.} \label{fig1}
\end{figure}

A critical input to the classical model is
the local projectile breakup probability ${\cal P}^L_{BU}(R)$ at a
separation $R$ from the target, deduced from ${P}_{BU} (R_{min})$.
To allow comparison with the CDCC, ${P}_{BU} (R_{min})$ has been deduced
from the CDCC breakup cross sections, $\sigma_{J}(\mbox{\sc ncbu})$,
for each $P$-$T$ partial wave. This is well represented
by an exponential function (beyond the strongly absorbed
partial waves), with $A=1.38$ and $\alpha=0.293$ fm$^{-1}$ 
[see Eq.\ (\ref{one})]. Experimental parameters defining ${P}_{BU} (R_{min})$ 
have been deduced from breakup measurements for the reaction 
$^{9}$Be+$^{208}$Pb for a wide range of sub-barrier energies \cite{HindeBU}. 
The parameters which best describe the experimental 
{\sc ncbu} yields are $A=1.68\times 10^4$ and $\alpha=0.922$ fm$^{-1}$. 
The CDCC (dashed line) and empirical ${P}_{BU}(R_{min})$ 
(thick solid line) distributions are compared in Fig.\ \ref{fig1}.
It is significant that the two functions are very different, the slope 
parameter $\alpha$ depending sensitively 
on the reaction and the breakup mechanism. For instance, our model of 
$^{8}$Be breakup, unlike the physical $^{9}$Be, does not generate 
Coulomb or nuclear dipole breakup couplings. Nor does it include 
the effect of neutron 
transfer channels, that may be significant in the empirical distribution 
\cite{Alexis2}. This comparison shows that
the ${\cal P}^L_{BU}(R)$ have high diagnostic value, and many reaction 
observables are also sensitive to the assumed ${\cal P}^L_{BU}(R)$, 
as shown below.

We can now use the CDCC-derived ${\cal P}^L_{BU}(R)$ as the sampling
distribution in the stochastic classical model calculations to
assess the consistency of the observables calculated with the quantum
mechanical and classical dynamical models. In sampling breakup 
$\ell_{max}$ = 4$\hbar$ and $\varepsilon_{max}$ = 6 MeV. 
Guided by the calculated g.s. 
probability of the projectile, the Gaussian distribution of $d_{12}$ 
values is taken to have a centroid and variance of 1.8 fm and 1.0 fm, 
respectively. The same real potentials are used
as in the CDCC when solving the post-breakup three-body problem. The
potential used for the $P$($^8$Be)-$T$ system prior to breakup, (120.9
MeV, 1.39 fm, 0.76 fm), is consistent with the folding model potential 
derived from the $\alpha$--$T$ potentials and the g.s. wave function 
of $P$. In this way the extended size and weak binding of $P$ are 
taken into account prior to breakup. Values of $L_0 \leq 100\hbar$ 
were found to be sufficient for the convergence of the observables.

\begin{figure}
\begin{tabular}{cc}
\includegraphics[width=0.48\textwidth,angle=0]{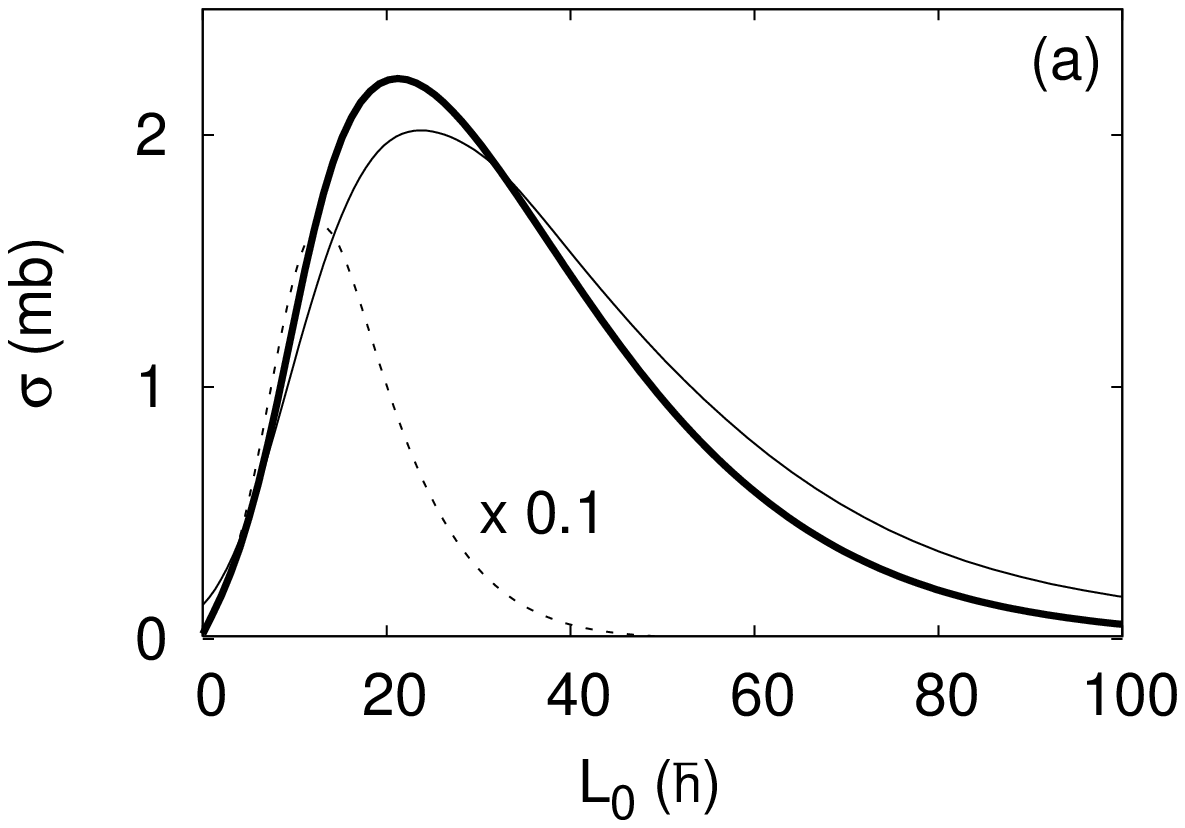} &
\includegraphics[width=0.48\textwidth,angle=0]{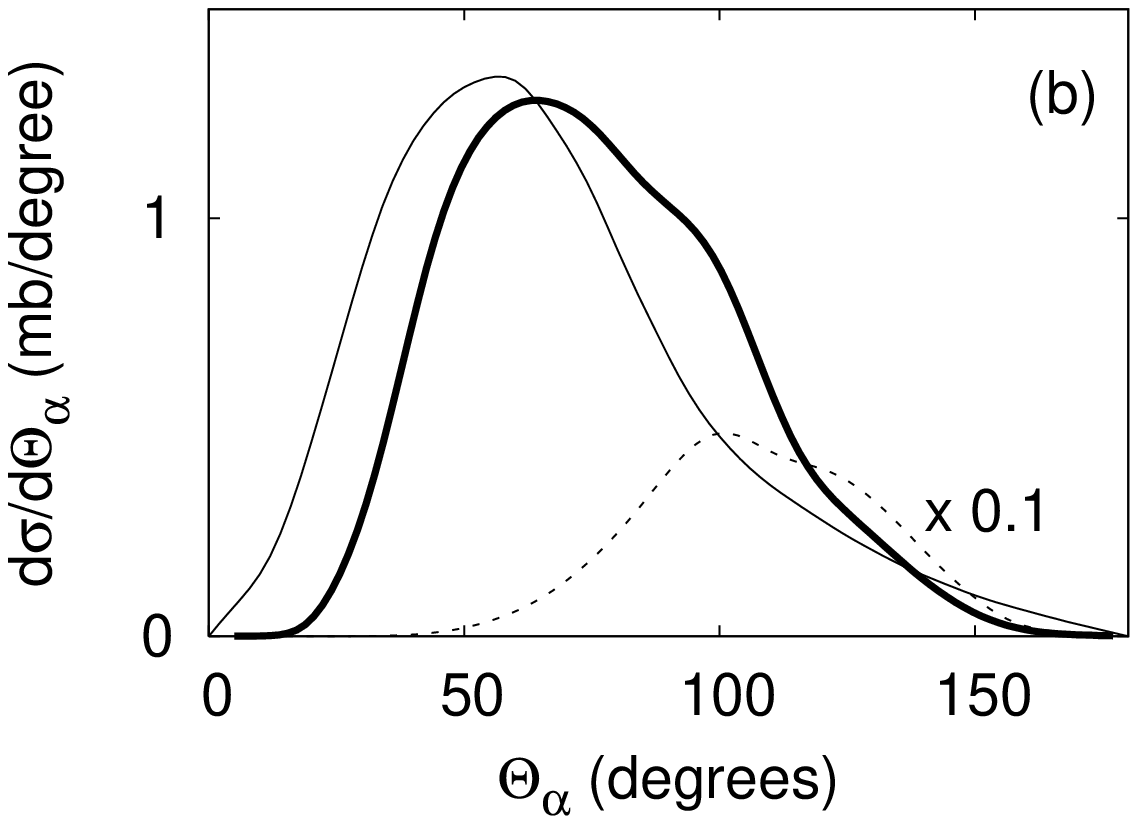} \\
\includegraphics[width=0.48\textwidth,angle=0]{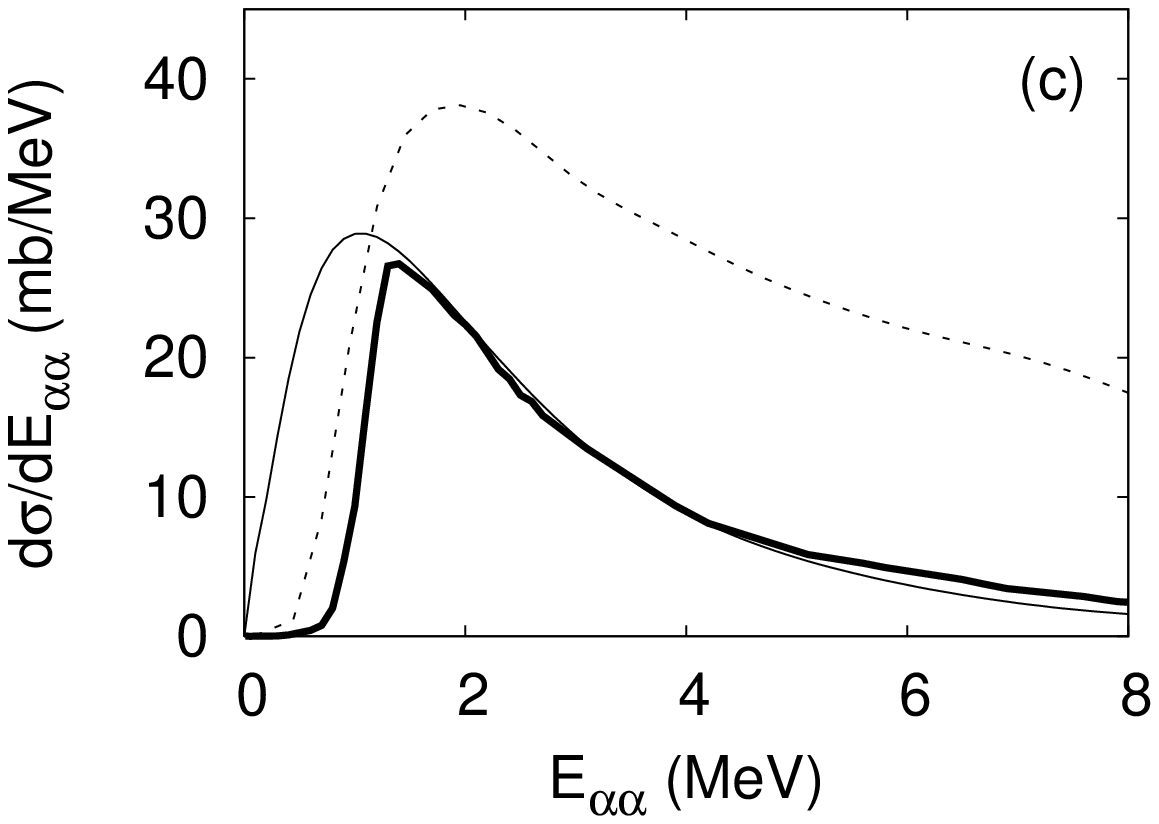} &
\includegraphics[width=0.48\textwidth,angle=0]{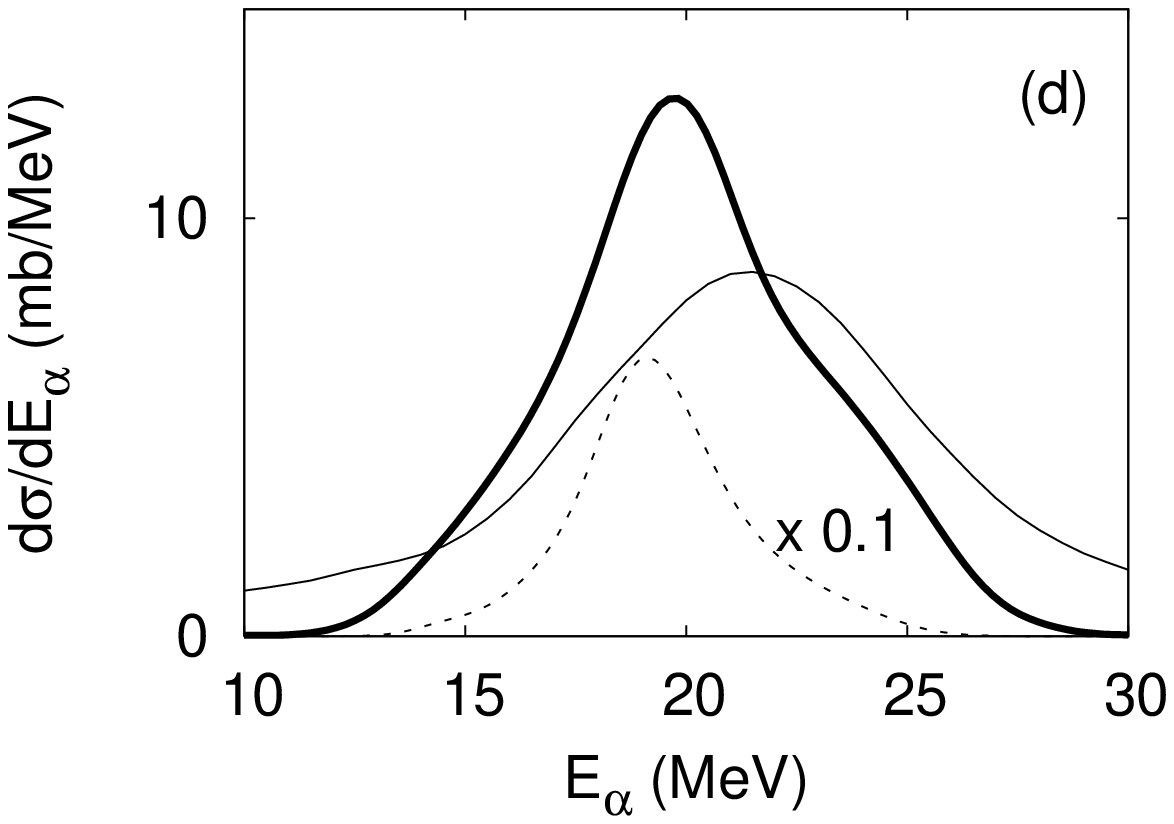}
\end{tabular}
\caption{The fully quantum mechanical CDCC model {\sc ncbu} observables 
are shown by the thin solid curves. Those from the classical model using
the same breakup probability function are quite similar (thick solid curves). 
Using an empirical ${\cal P}^L_{BU}(R)$ deduced from the $^{9}$Be+$
^{208}$Pb reaction gives very different results (dashed curves), which are 
scaled by 0.1 where noted, demonstrating the sensitivity of the observables 
to ${\cal P}^L_{BU}(R)$.} \label{fig3}
\end{figure}

The calculated total absorption and total {\sc ncbu} cross sections
from the CDCC are $\sigma({\rm abs})$=327 mb and $\sigma({\mbox{\sc
ncbu}})$=97 mb, compared with $\sigma({\rm abs})$=300 mb and
$\sigma({\mbox{\sc ncbu}})$=93 mb from the classical model. 
Fig. \ref{fig3} compares the CDCC (thin solid curves) and the
classical model (thick solid curves) for (a) the breakup cross
sections as a function of $P$-$T$ partial wave, (b) the singles
$\alpha$-particle angular distribution, (c) the $\alpha$-$\alpha$
relative energy distribution, and (d) the singles $\alpha$-particle
energy distribution. The classical and CDCC results 
are seen to agree quite well, both qualitatively and quantitatively, 
providing support for the classical dynamical treatment.

Fig. \ref{fig3} also shows the strong sensitivity of the (classical
model) observables to the breakup function assumed. The dashed curves 
result from using the empirical ${\cal P}^L_{BU}(R)$ from the $^{9}$Be+$^{208}$Pb 
experiment, as opposed to the CDCC breakup function 
(thick solid curve). The former 
results in breakup occuring much nearer to $R_{min}$, and thus to the target.
The {\sc ncbu} spin distribution when using 
the empirical breakup function (panel a) now peaks at smaller angular 
momenta with a corresponding shift in the peak of the $\alpha$ angular 
distribution to larger angles (panel b). 
The strong tidal forces experienced by the non-fused $\alpha$-particles 
then lead to an enhanced $\alpha$-$\alpha$ yield of high relative energy 
(panel c) and corresponding changes to the singles $\alpha$ energies 
(panel d). 
 
These first results already 
indicate a potentially very useful, and very sensitive mapping 
between (the unobservable) ${\cal P}^L_{BU}(R)$ and the {\sc ncbu} observables 
shown in Fig. \ref{fig3}. The breakup function, and reaction dynamics,
probed by combining the classical model and experimental data, could thus provide 
much needed insight into the localisation and nature of the breakup and fusion 
mechanism for different reacting systems.

\begin{figure}
\includegraphics[width=0.65\textwidth]{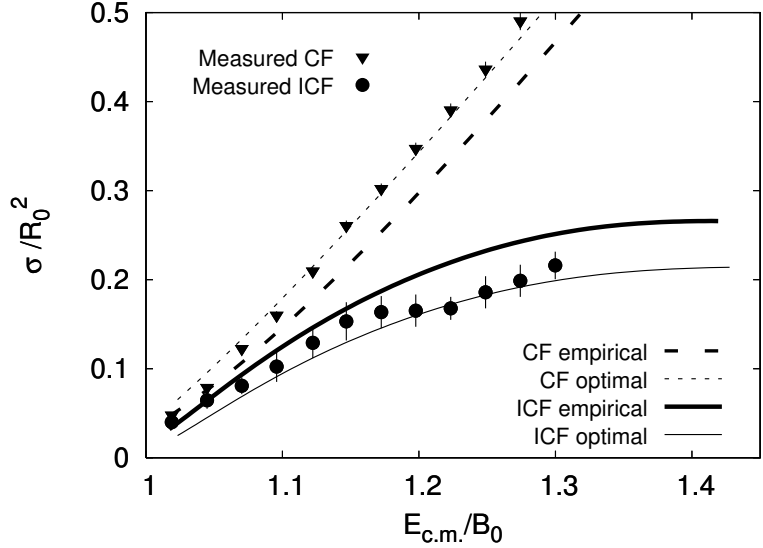}
\caption{The classical model {\sc cf} (dashed curves) and {\sc icf} 
(solid curves) cross sections for $^{8}$Be+$^{208}$Pb 
are compared with the $^{9}$Be+$^{208}$Pb experimental 
data of Ref.\ \protect\cite{Dasgupta0}. 
The energy and the cross section of the calculated and experimental data are 
normalized by the barrier energies ($B_0=40.7$ MeV for thick curves, $B_0=40.5$ MeV for thin curves, 
and $B_0=37.5$ MeV for experimental data) 
and their respective radii ($R_0=10.7$ fm for theoretical curves, and $R_0=10.8$ fm for experimental data), 
obtained by fitting the above-barrier total fusion ({\sc cf}+{\sc icf}) cross sections 
with the formula $\sigma_{fus} = \pi R_{0}^{2}(1-B_{0}/E_{c.m.})$.} 
\label{fig4}
\end{figure}

Finally, we apply these ideas to experimental data. Using ${\cal P}^L_{BU} (R)$ deduced from the empirical 
breakup function (thick solid line in Fig. \ref{fig1}), the model predicts {\sc cf} and 
{\sc icf} yields as shown by the thick solid and dashed curves in Fig. \ref{fig4}. 
These are close to the experimental data (triangles and circles). Only a small change in ${\cal P}^L_{BU} (R)$ 
is required to give a very good reproduction of the data (thin lines in Figs. \ref{fig1} and \ref{fig4}). 
This result is the first quantitative result supporting the suggested linkage \cite{HindeBU} between the 
breakup process observed below the barrier and suppression of complete fusion observed at energies above 
the barrier. 
 
In summary, a three-dimensional classical dynamical model for low
energy reactions of weakly-bound nuclei is presented, which for the first time 
allows consistent realistic calculations relating below-barrier {\sc ncbu} with 
above-barrier {\sc cf} and {\sc icf}. 
In the test case studied, the model, which treats the break-up process
stochastically, is shown to reproduce the differential {\sc ncbu}
observables of the CDCC quantum approach quite accurately. We have
shown that the breakup probability function ${\cal P}^L_{BU}(R)$ 
has high diagnostic value, being sensitive to the reaction mechanism
and reactants, and that {\sc ncbu} reaction observables are strongly 
sensitive to and can interrogate this model input. The model also 
demonstrates the importance of the breakup probability function, 
accessible from experiments at sub-barrier energies, to predictions 
of complete and incomplete fusion yields of weakly-bound projectiles 
at energies above the Coulomb barrier.

All authors acknowledge support from an ARC Discovery grant whilst JAT 
acknowledges support also from 
the United Kingdom Engineering and Physical Sciences Research
Council (EPSRC) under Grant No. EP/D003628.

\end{document}